\documentclass[journal]{IEEEtran}
\usepackage{times}
\usepackage{epsfig}
\usepackage{graphicx}
\usepackage{amsmath}
\usepackage{amssymb}
\usepackage{algorithm}
\usepackage{algpseudocode}
\ifCLASSINFOpdf
\else
\fi
\begin{document}

%
\title{Low Dose CT Denoising via Joint Bilateral Filtering and Intelligent Parameter Optimization}

\author{Mayank~Patwari,
	Ralf~Gutjahr,
	Rainer~Raupach,
	and~Andreas~Maier
	\thanks{M. Patwari, and A. Maier are with the Pattern Recognition Lab at the Friedrich-Alexander University of Erlangen-Nuremberg, 91058, Erlangen, Germany e-mail: mayank.patwari@fau.de.}
	\thanks{M. Patwari, R. Gutjahr, and R. Raupach are with Siemens Healthcare GmbH, 91301, Forchheim, Germany.}
}

%


\maketitle

\begin{abstract}
Denoising of clinical CT images is an active area for deep learning research. Current clinically approved methods use iterative reconstruction methods to reduce the noise in CT images. Iterative reconstruction techniques require multiple forward and backward projections, which are time-consuming and computationally expensive. Recently, deep learning methods have been successfully used to denoise CT images. However, conventional deep learning methods suffer from the 'black box' problem. They have low accountability, which is necessary for use in clinical imaging situations. In this paper, we use a Joint Bilateral Filter (JBF) to denoise our CT images. The guidance image of the JBF is estimated using a deep residual convolutional neural network (CNN). The range smoothing and spatial smoothing parameters of the JBF are tuned by a deep reinforcement learning task. Our actor first chooses a parameter, and subsequently chooses an action to tune the value of the parameter. A reward network is designed to direct the reinforcement learning task. Our denoising method demonstrates good denoising performance, while retaining structural information. Our method significantly outperforms state of the art deep neural networks. Moreover, our method has only two parameters, which makes it significantly more interpretable and reduces the 'black box' problem. We experimentally measure the impact of our intelligent parameter optimization and our reward network. Our studies show that our current setup yields the best results in terms of structural preservation.
\end{abstract}


%
\IEEEpeerreviewmaketitle

\section{Introduction}

\noindent The radiation dose in clinical CT needs to be as low as reasonably achievable to warrant patient's safety. Reducing radiation dose creates noise in the reconstructed images, which may obscure critical details and low contrast objects \cite{Oppelt2005a}. The non-stationary nature of CT noise means that simple filtering approaches are not always successful in denoising. Moreover, filtering also removes the edge information present in the image, particularly the small low-contrast features . Therefore, denoising the reconstructed images requires a more intelligent solution, and is currently an active area of research.

Non linear edge-preserving filters such as the bilateral filter have been used to denoise clinical CT data \cite{Manhart2014, Manduca2009}. Most clinically approved methods currently use iterative reconstruction methods \cite{Gilbert1972} to denoise images \cite{Ramirez-Giraldo2015, Angel2012}. Deep learning has also been successfully applied to the clinical CT denoising problem \cite{ Wolterink2017, Shan2018, Maier2015}. However, iterative reconstruction methods are time-consuming and computationally expensive. Deep learning methods, while outperforming iterative reconstruction, suffer from a lack of accountability, the so-called 'black box' problem. It is absolutely crucial for any denoising approach to be completely interpretable for possible clinical approval. A middle ground between the underperforming known methods and deep learning would be to use deep learning to tune or optimize the performance of known techniques, such as by optimizing the parameters \cite{Zhu2010} or altering the regularization strength \cite{Shen2018}.

\begin{figure}[t]
	\centering
		\includegraphics[width=0.93\linewidth]{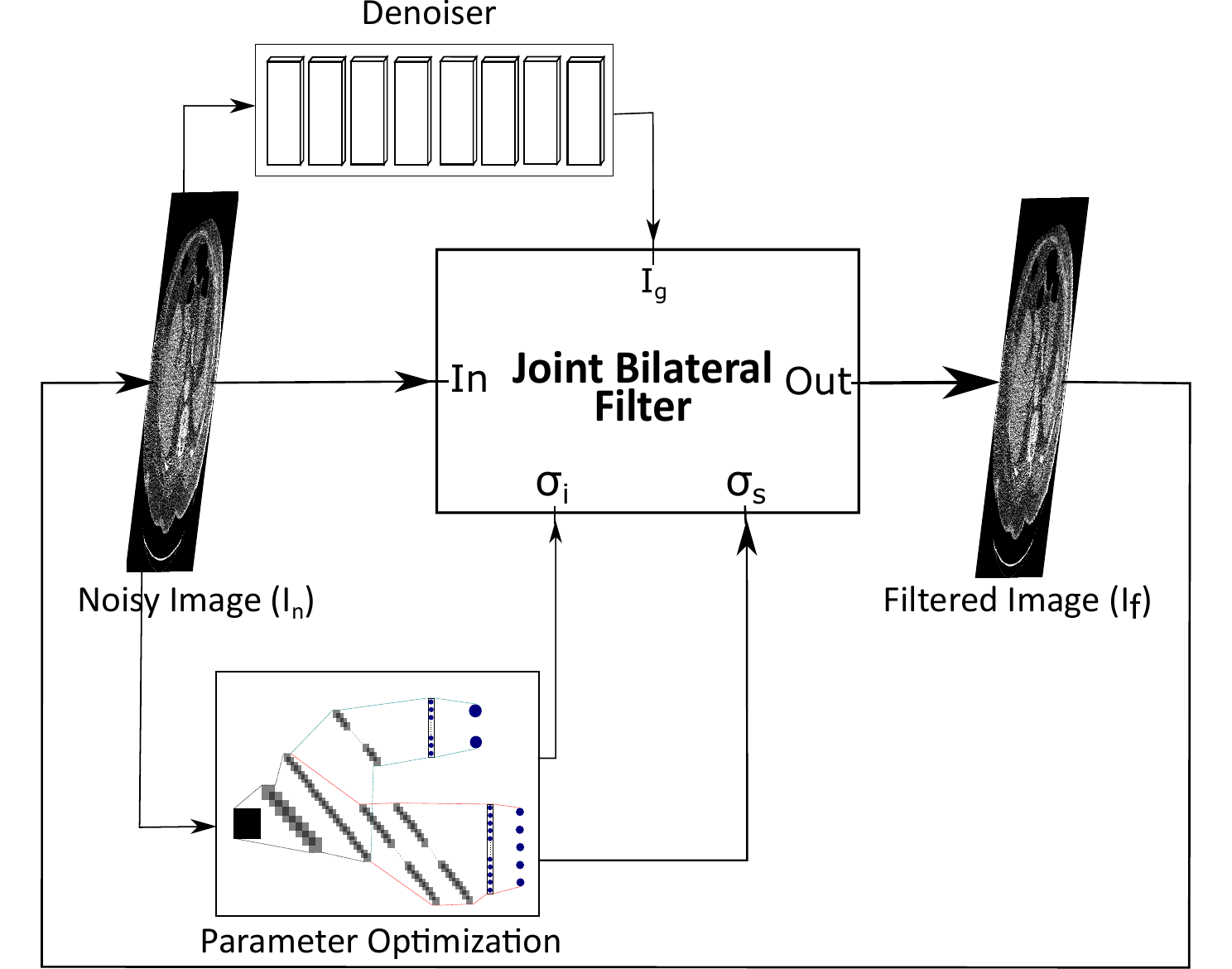}
		\caption{Our denoising scheme with the JBF, parameter optimization scheme, and guidance image estimating denoiser.}
		\label{denoisemethodfig}
\end{figure}

In this paper, we aim to combine the strengths of both known operators and deep learning. We introduce a Joint Bilateral Filter (JBF) \cite{Manhart2014} with a deep learned guidance image to denoise the image. The JBF has two parameters, which are tuned by reinforcement learning \cite{Sutton2014}. The reinforcement learning scheme tunes our parameters pixelwise to obtain the best quality image overall. The JBF is applied iteratively for a given number of iterations. The parameters are optimized for every iteration. We compare our methods to state of the art deep learning networks for the task of low dose CT denoising. Our proposed method shows good denoising performance and preserves structural information. Furthermore, our method outperforms deep learning methods, while being significantly more interpretable.

\section{Materials and Methods}

\subsection{Joint Bilateral Filtering concept}
\noindent We revisit the concept of the JBF for better explaining our methods. The JBF is a variant of the standard bilateral filter, which uses a separate guiding image to smooth the difference in intensities. The operation of the JBF is defined by the following equation:
\begin{equation}
\label{JBFeqn}
I_{f} (x)=  \frac{\sum_{o\epsilon N(x)} I_n(o)G_{\sigma_s}(x-o)  G_{\sigma_i}(I_g (x)- I_g (o))}{\sum_{o\epsilon N(x)} G_{\sigma_s}(x-o)  G_{\sigma_i}(I_g (x)- I_g (o))}
\end{equation} 
where $I_n$ is the noisy image, $I_f$ is the filtered image, $x$ is the spatial coordinate, $N(x)$ is the neighborhood of $x$ taken into account for the filtering operation, $I_g$ is the guidance image estimated by the method described in Section \ref{JBFprior} and $G_\sigma$ is a Gaussian operator. The first Gaussian operator in Equation \ref{JBFeqn} is a spatial filter, while the second operator is the range filter. The Gaussian operator is defined by 
\begin{equation}
G_\sigma(x)=  \frac{e^{-x^2 / 2\sigma^2 }}{2\sigma^2}
\end{equation}
where $\sigma$ is a parameter controlling the level of smoothing. $\sigma$ is usually a hand tuned parameter. Therefore, the JBF at any point has three inputs, the guidance image $I_g$, and two $\sigma$ parameters, $\sigma_i$ for the range operator, and $\sigma_s$ for the spatial operator.  We used a Gaussian operator of size 9 $\times$ 9 $\times$ 5 across the entire image. 

\subsection{Quality Analysis}
\label{qualityCNN}

\begin{figure}[t]
	\includegraphics[width=\linewidth]{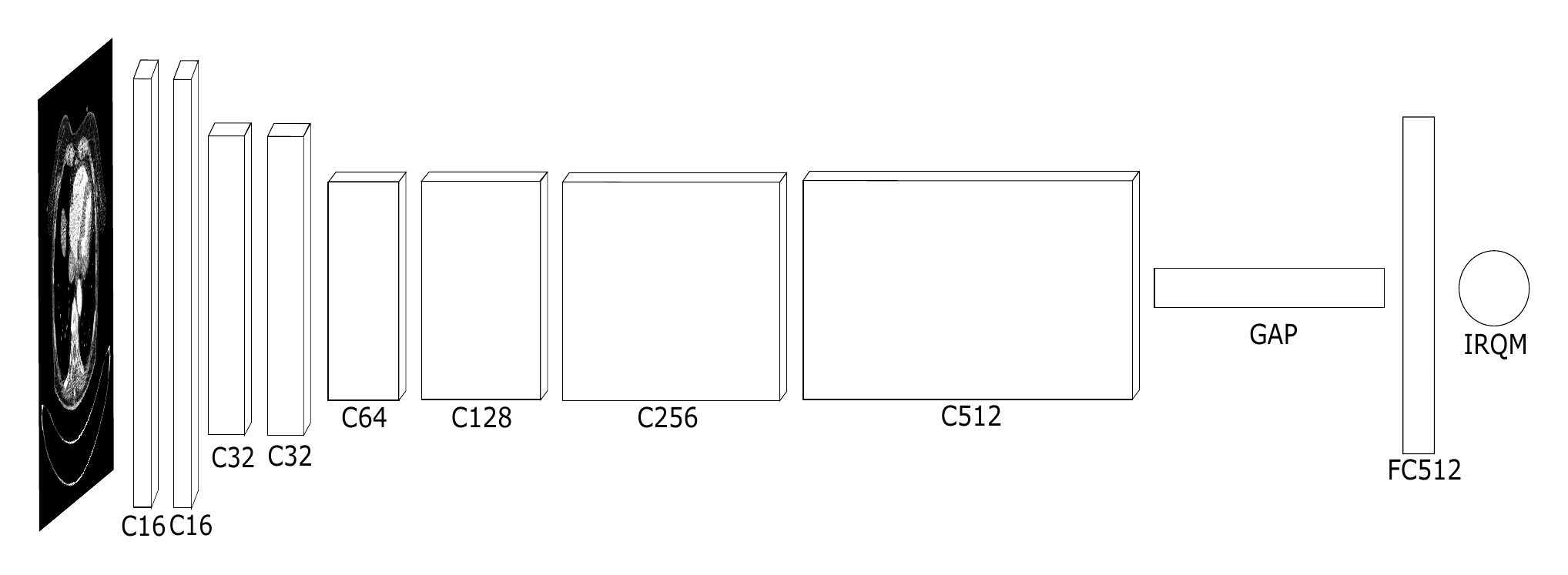}
	\caption{The quality network used to measure image quality and noise. C represents the number of convolutional filters.}
	\label{qualitynetfig}
\end{figure}

\noindent We train a CNN to estimate the Image Reconstruction Quality Metric (IRQM). We use the architecture and training hyperparameters described in Patwari et al. \cite{Patwari2019}, with two differences \textbf{(1)} we leave out the batch normalisation layers \textbf{(2)} the output of the final neuron is passed through a sigmoid layer to constrain the output to [0, 1]. The network is trained using 4 CT volumes. Each volume was reconstructed with standard clinical dose, and with doses reduced by 50\% and 75\%. The gradient structural similarity score between the standard dose image and the reduced dose image is used as a proxy quality score for training. The heatmap loss is included as a secondary loss function. A network diagram is present in Fig. \ref{qualitynetfig}.

\subsection{Estimating a Guidance Image}
\label{JBFprior}
 
\begin{figure}[b]
	\includegraphics[width=\linewidth]{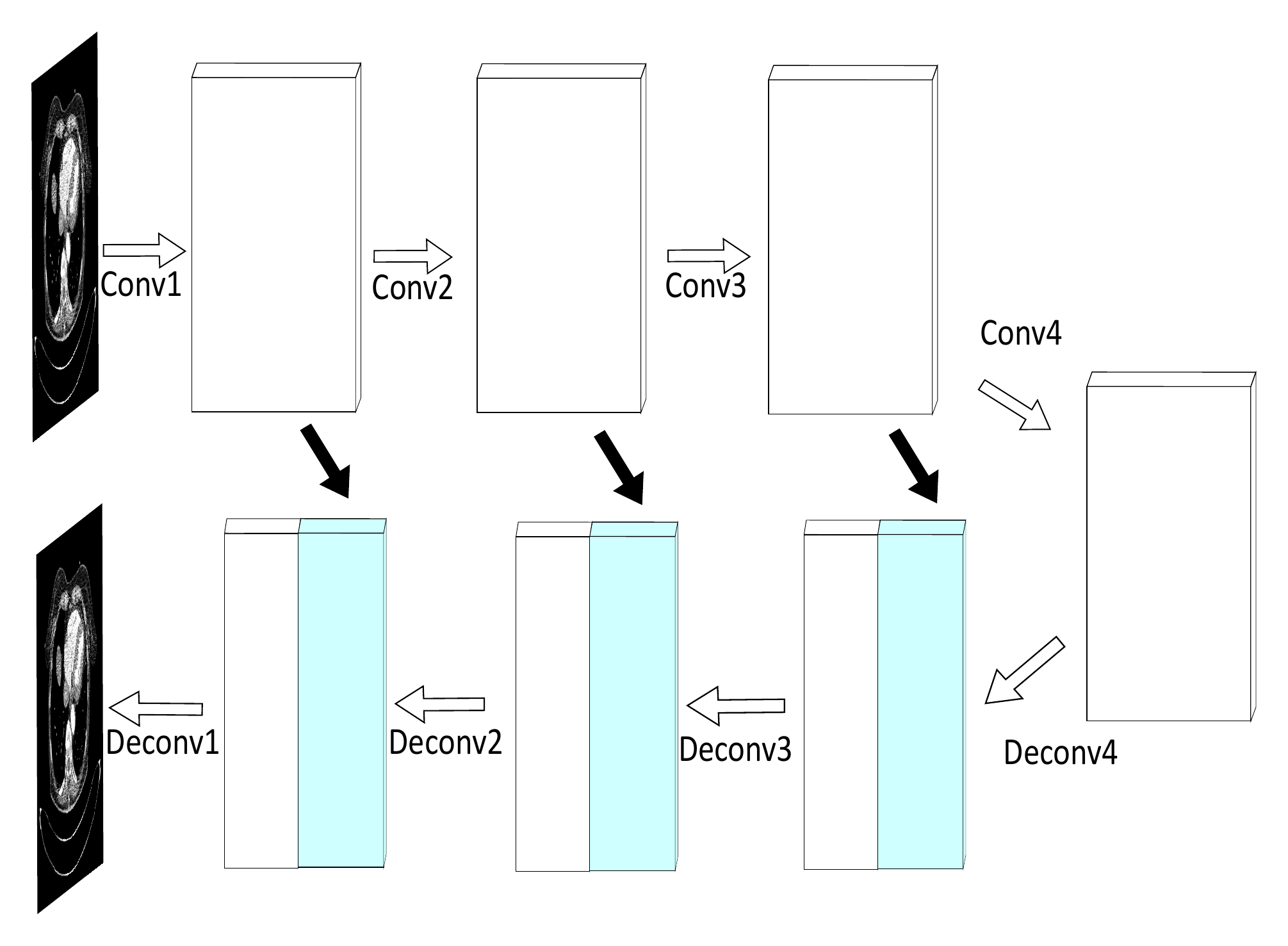}
	\caption{The residual network used to create the guidance image. Black arrows represent the residual connections.}
	\label{denoisenetfig}
\end{figure}

 \noindent The guidance image is estimated to be a noise free image. We can solve the following optimization problem to find a noise removal function $F$:
\begin{equation}
\label{minREDCNN}
\min_F ||I_{nf} - F(I)||^2
\end{equation}
where $I_{nf}$ is the noise free image and $I$ is the noisy image. Solving equation \ref{minREDCNN} will give us the ideal noise removal function $F$.
We parameterize $F$ using a residual CNN (Fig. \ref{denoisenetfig}) \cite{Shan2018}. The residual CNN has 4 convolutional layers and 4 deconvolutional layers, each with 32 filters. The convolutional layers have filters of size 3 $\times$ 3 $\times$ 3. No padding is applied, shrinking the input image in each dimension. The deconvolutional filters have filters of size 3 $\times$ 3. This results in a receptive field of 17 $\times$ 17 $\times$ 9.  A leaky ReLU activation follows each layer, except the final layer.  

 The network is trained using CT volumes of 10 patients. Each volume is reconstructed with the standard clinical dose and with simulated dose reductions. The standard dose images act as ground truth volumes $I_{nf}$. The reduced dose volumes are reconstructed at doses of 5\%, 10\%, 25\% and 50\% of the standard clinical dose, and are used as training volumes. The network is trained using non-overlapping image patches of size 64 $\times$ 64 $\times$ 9. This is shrunk to 56 $\times$ 56 $\times$ 1 by the convolutional layers, and expanded to 64 $\times$ 64 by the deconvolutional layers. The training took place over 70,000 iterations with 48 patches per iteration.

\subsection{Tuning parameters via deep reinforcement learning}

\noindent We introduce a convolutional network $NET$ as our agent, which would choose a parameter and subsequently choose the action that affects this parameter. We limit the network to choose between the two $\sigma$ parameters, and between the following five actions \textbf{(1)} increase by 50\% \textbf{(2)} increase by 10\% \textbf{(3)} do nothing \textbf{(4)} decrease by 10\% \textbf{(5)} decrease by 50\%.

The network is used to maximize a reward function $r$.  $r$ in this case is the output of our quality network $Q_n$, described in section \ref{qualityCNN}. 
\begin{equation}
r = Q_n(s') - Q_n(s)
\end{equation}
where $s'$ is the image after the parameter tuning step and $s$ is the original image. We create a loss which punishes deviations from the Bellman equation. The Bellman equation is given by
\begin{equation}
Q^*(s, a) = r + \gamma \max_{a'}Q^*(s', a')
\end{equation}
where $a$ is the action which changes $s$ to $s'$ and $\gamma$ is the discount factor for future rewards.

\begin{figure}[b]
	\includegraphics[width=\linewidth]{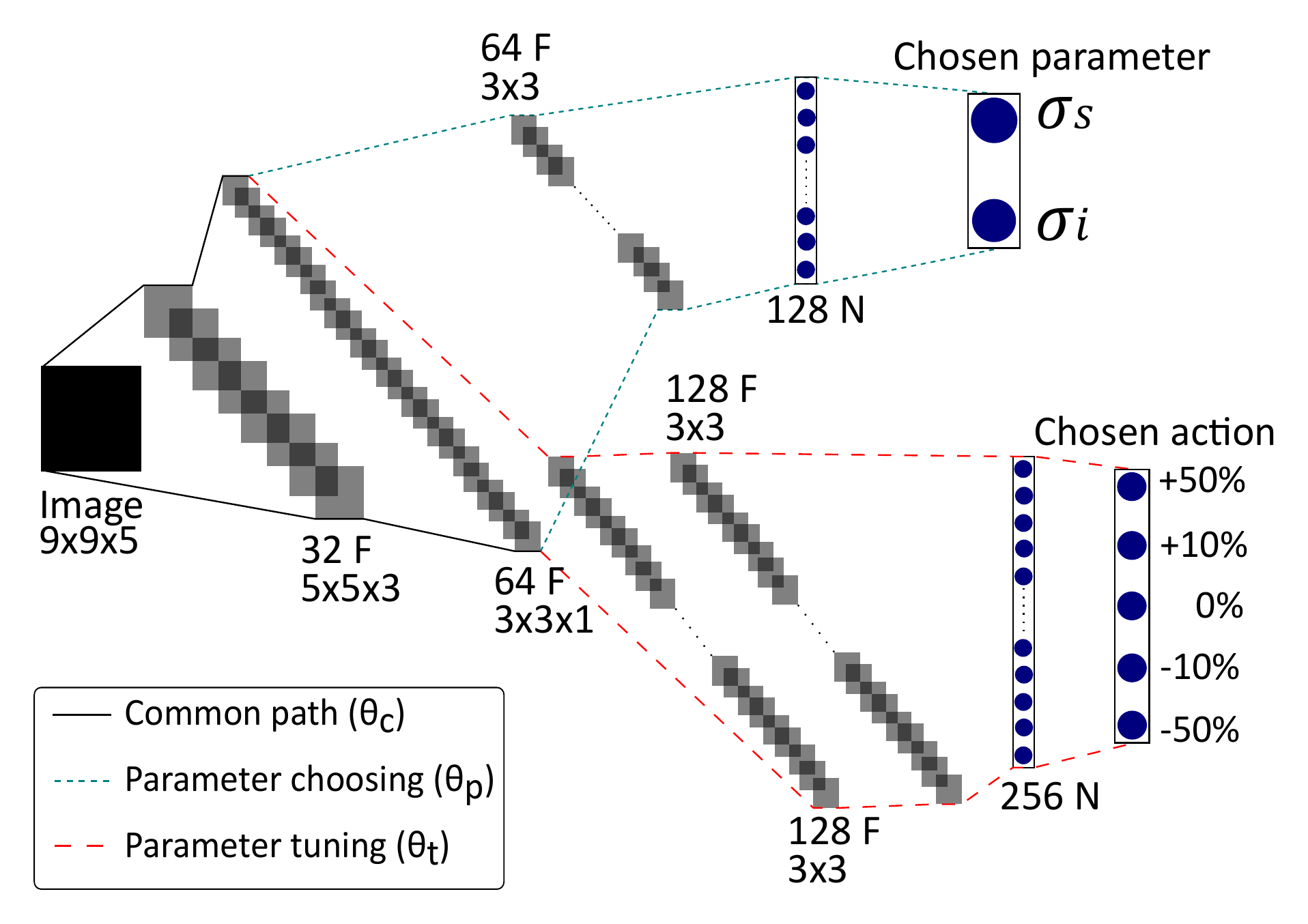}
	\caption{Our RL network architecture with separate parameter choosing and parameter tuning paths. F represents the number of convolutional filters, N the number of neurons, and the dimensions the size of the outputs.}
	\label{rlnetfig}
\end{figure}

We parameterize the $Q^*$ function with our CNN $NET$. $NET$ has three sets of parameters, $\theta_p$ which influence the parameter choosing, $\theta_t$ which control the parameter tuning, and $\theta_c$, which are shared parameters. We split our target term into two target terms. Following standard reinforcement learning techniques \cite{Sutton2014}, we use an older version of our policy network ($NET’$) as a target.
\begin{equation}
\begin{split}
y_1 = r + \gamma \max_{p'}Q^*(s', p'; NET'(\theta_c, \theta_p)) \\
y_2 = r + \gamma \max_{t'}Q^*(s', t'; NET'(\theta_c, \theta_t))
\end{split}
\end{equation}
The network is trained by solving the following minimization problem:
\begin{equation}
\label{losseqn}
\begin{aligned}
&L_1 = {[y_1 - Q(s, p; NET(\theta_c, \theta_p))]}^2 \\
&L_2 = {[y_2 - Q(s, t; NET(\theta_c, \theta_t))]}^2 \\
&\min_{NET} L = L_1 + L_2 \\
\end{aligned}
\end{equation}

The network chosen (Fig. \ref{rlnetfig}) has two paths, one which chooses the parameter to be tuned, and the other which tunes the parameters. The first path has 3 convolutional layers, one fully connected layer, and an output layer. The convolutional layers have 32, 64, and 64 layers. The fully connected layer has 128 neurons. There are two output neurons. The second path has 4 convolutional layers, one fully connected layer, and one output layer. The convolutional layers have 32, 64, 128 and 128 layers. The fully connected layer has 256 neurons. There are 5 output neurons. The first two convolutional layers are shared between the paths, and have 3 $\times$ 3 $\times$ 3 convolutions. All other convolutional layers have 3 $\times$ 3 convolutions. All layers except the output layers are followed by leaky ReLU activations.

The network is trained for 120,000 iterations. $\gamma$ is fixed at 0.99. $NET'$ is updated to $NET$ every 300 steps. We allow 20 tuning steps before application of the bilateral filter. The input to the network is an image patch of size 9 $\times$ 9 $\times$ 5.  The training set consists of 20 slabs of size 256 $\times$ 256 $\times$ 16 reconstructed at 25\% dose. The parameter chosen, action chosen, slab, filtered slab, and increased quality score are calculated for each patch. 3200 patches are added at random into the training pool. 512 patches are chosen at random for training at each optimization step.


\section{Experiments and Results}

\begin{figure*}
	\includegraphics[width=\linewidth]{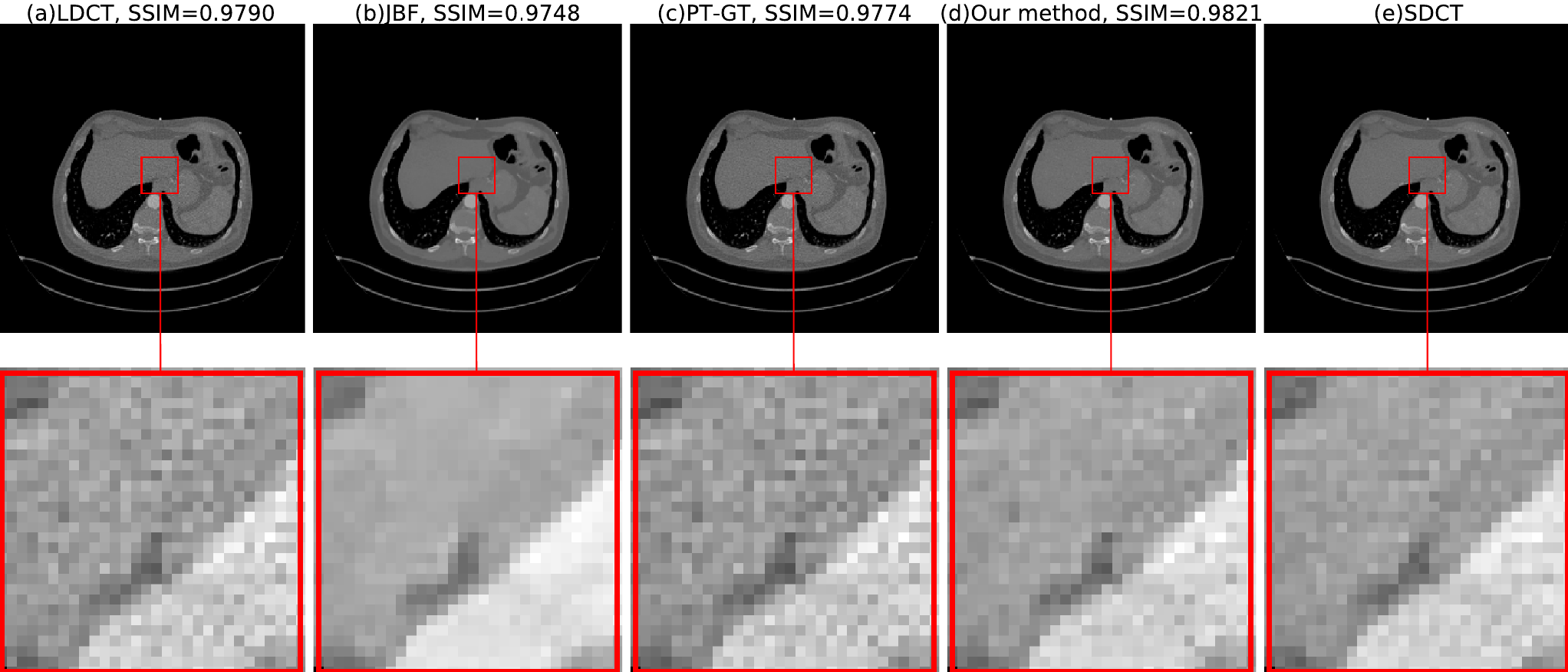}
	\caption{An image showing the (a) reduced dose CT image (b) image denoised by JBF (c) image denoised by using the reinforcement learning technique which optimizes based on the ground truth image (PT-GT) (d) our method and (e) the standard dose reference image. Our method yields the closest looking image to the reference image (SSIM = 0.9821). The images are reconstructed with a window of [-800, 1200].}
	\label{denoisefig}
\end{figure*}

\begin{table}[t]
	\caption{Mean and standard deviations of the quality metric scores of our method and variants. Plain JBF application yields the highest PSNR, while including network based parameter tuning yields the highest SSIM.}
	\label{resulttable}
	\begin{center}
		\begin{tabular}{|c||c | c| }
			\hline
			& \textbf{PSNR} & \textbf{SSIM}\\
			\hline
			Low Dose CT \cite{Mccollough2016} & $43.12 \pm 1.021$ & $0.9636 \pm 0.0073$  \\
			\hline
			Joint Bilateral Filter (JBF) & $\boldsymbol{46.79 \pm 0.8685}$ & $0.9770 \pm 0.0046$ \\
			JBF + PT-GT & $43.40 \pm 1.025$ & $0.9656 \pm 0.0072$ \\
			\hline
			CPCE3D \cite{Shan2018} & $45.43 \pm 0.6914$ & $0.9817 \pm 0.0029$ \\
			GAN\cite{Wolterink2017} & $41.87 \pm 0.7079$ & $0.9398 \pm 0.0079$ \\
			\hline
			\textbf{Our method}  & $45.02 \pm 0.9709$ & $\boldsymbol{0.9856 \pm 0.0027}$ \\
			\hline
		\end{tabular}		
	\end{center}
\end{table}

\subsection{Data and Evaluation Metrics}
We used the data of the Grand Low Dose CT challenge \cite{Mccollough2016} as our testing data. This dataset consists of body CT scans of 10 different patients, each reconstructed with both standard dose, and with 25\% of the standard dose. We use the images reconstructed with 1mm slice thickness with a medium smooth B30 kernel. The reduced dose images are our input images, while the standard dose images are treated as reference images.

We use the PSNR and SSIM \cite{Wang2004} as our metrics for noise removal and structural preservation respectively. The SSIM is a significantly more important predictor of the diagnostic quality of the image, as the PSNR promotes smoothing of images. Due to the limited number of patients, we use the Wilcoxon's signed rank test to assess statistical significance. A p-value of less than 0.05 is considered as the threshold for statistical significance.

\subsection{Denoising and comparison to state of the art}
In our denoising experiments, we denoise the image by iterative application of the JBF. For each iteration, we allow five optimization steps to tune our parameters, before applying the bilateral filters. The parameters are reset to their original guesses and the optimization process starts anew for each iteration. We restrict the number of JBF iterations to four in our testing phase.

Our method significantly improved PSNR and SSIM compared to the noisy image (w = 0.000, p = 0.0051). This shows that our method reduces image noise while preserving the structural information present in the images. An example can be seen in Fig. \ref{denoisefig}. Our method outperforms state of the art deep learning networks such as CPCE3D \cite{Shan2018} and GAN \cite{Wolterink2017} in terms of SSIM. However, CPCE3D has a higher PSNR than our method. This could indicate superior denoising at the cost of feature preservation. The results can be viewed in Table \ref{resulttable}.

\subsection{Ablation study}
\textbf{Impact of parameter tuning:} We observe the impact of the JBF with and without parameter tuning. Use of parameter tuning improves the SSIM (w = 0.000, p = 0.0051), but significantly decreases the PSNR (w = 0.000, p = 0.0051). Therefore, the use of parameter tuning reduces the amount of denoising, but improves the amount of structural information preserved.

\textbf{Using our reward network:} We compare the use of our reward network (Section \ref{qualityCNN}) with the mean-squared-error based reward function used in Shen et. al. \cite{Shen2018}. We call this method parameter tuning - ground truth (PT-GT). PT-GT reduces both the PSNR and the SSIM significantly (w = 0.000, p = 0.0051). Therefore, the use of our reward network results in better CT image quality compared to PT-GT.

\section{Conclusion}
In this study, we introduce a method which tunes JBF parameters to thoroughly denoise low dose CT images. Our method provides good denoising performance while retaining the structural information. The use of the JBF reduces noise, while the parameter optimization helps in retaining structures.  Our method is outperformed several existing state of the art solutions for low dose CT denoising. Additionally, our method is significantly more interpretable, and reduces the ‘black box’ problem that exists in deep learning, which improves applicability in clinical imaging.





\bibliographystyle{IEEEtran}
%
%
%
\bibliography{Bibliography}

\end{document}